\documentclass[10pt, aps,pra,twocolumn,superscriptaddress,longbibliography,nofootinbib]{revtex4-2}
\usepackage{amsmath, amsfonts, amsthm, amssymb}
\usepackage{graphicx,xcolor,tikz}
\usepackage{hyperref}
\usepackage{physics}

\usepackage{dcolumn}
\usepackage{bm}

\newcommand{\pr}{\mathrm{pr}}
\newcommand{\hn}{\widetilde{H}}

\begin{document}

\title{Entropy threshold: A simple proxy for performance of quantum error correction}

\author{Diego Ruiz}
\thanks{Current affiliation: AWS Center for Quantum Computing, Pasadena, CA, USA}
\affiliation{Yale Quantum Institute \& Department of Applied Physics, Yale University, New Haven, CT, USA}
\affiliation{Alice \& Bob, 49 Bd du G\'en\'eral Martial Valin, 75015 Paris, France}
\affiliation{Laboratoire de Physique de l'\'Ecole Normale Sup\'erieure,
\'Ecole Normale Sup\'erieure, Centre Automatique et Syst\`emes,
Mines Paris, Universit\'e PSL, CNRS, Inria, Paris, France}
\author{Aleksander Kubica}
\affiliation{Yale Quantum Institute \& Department of Applied Physics, Yale University, New Haven, CT, USA}

\date{\today}

\begin{abstract}
The rapidly growing landscape of quantum error-correction (QEC) protocols has produced a wealth of numerical data, but comparatively few heuristics for understanding and predicting their performance.
Here, we develop a simple entropy-based proxy that predicts the thresholds of a variety of QEC protocols, ranging from the code-capacity setting of Clifford-deformed surface codes with biased Pauli or erasure noise to the circuit-level noise model of the surface or color codes with flag qubits.
Our proxy estimates the threshold by locally comparing the noise entropy with the error information gained via stabilizer measurements (or spacetime detectors and flag outcomes in the circuit-level settings).
Despite neglecting correlations between stabilizer outcomes and the contribution from code degeneracy, the proxy captures the main trends across diverse settings and yields threshold estimates in good agreement with numerical results.
Our work develops much-needed phenomenology that enables simple back-of-the-envelope estimates of QEC performance, recovering and providing an explanation for the results of computationally intensive detailed simulations.
\end{abstract}

\maketitle

Quantum error correction (QEC) is essential for reliable, large-scale quantum computation~\cite{shor1995scheme, steane1996,shor1996fault,preskill1998}.
A central figure of merit for a QEC protocol is its threshold, which is the physical error rate below which the logical error rate can be made arbitrarily small by increasing the size of the underlying code~\cite{aharonov1997fault, knill1998resilient, aliferis2006quantum}.
For most QEC codes and noise models, however, thresholds are not known analytically, and estimating them requires computationally intensive Monte Carlo simulations of logical error rates across multiple code sizes and noise strengths.
The corresponding decoding tasks also face fundamental worst-case complexity barriers---minimum-weight decoding is NP-hard even for basic realizations of quantum memories and logical circuits~\cite{walters2026minimum,gu2026color}, while optimal decoding of stabilizer codes is \#P-hard~\cite{iyer2015hardness}.
These challenges motivate the search for simple proxies for comparing QEC protocols across different noise models.

In this article, we propose a simple heuristic to evaluate the threshold of a QEC protocol without the need for numerical simulations.
Our heuristic compares locally the noise entropy with the information about possible errors obtained via stabilizer measurements;
a threshold estimator is obtained by finding the roots of a single-variable equation, offering a quick performance test.
Although less precise than numerical simulations, our heuristic captures a wide range of trends across various QEC codes and noise models, including the code capacity setting for the surface, XZZX and XY codes under biased Pauli noise~\cite{bonilla2021xzzx, tuckett2018ultrahigh,dua2024clifford} or erasure noise with imperfect erasure checks~\cite{kubica2023erasure,wu2022erasure,gu2025optimizing,gu2025fault,chang2025surface,teoh2023dual,baranes2026leveraging}.
Furthermore, it extends to more realistic scenarios, including the surface code with the phenomenological or circuit-level noise models~\cite{dennis2002topological,fowler2012surface}, as well as the color code with or without flag qubits~\cite{chamberland2020triangular}.

\begin{figure*}
    \centering
    \includegraphics[width=\textwidth]{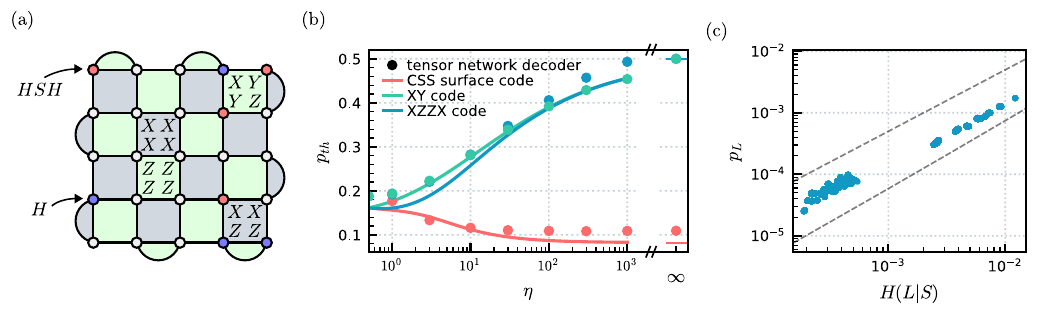} \caption{
    (a) Clifford-deformed surface codes~\cite{dua2024clifford}, including the XY and XZZX codes, are obtained from the surface code by applying $H$ and $HSH$ operators to some qubits (blue and red dots), modifying the original $X$ and $Z$ stabilizers (green and gray faces).
    (b) Threshold $p_{th}$ of the surface, XY and XZZX codes as a function of the noise bias $\eta$.
    Solid lines capture the entropy threshold; dots correspond to approximate maximum likelihood decoding with a tensor network decoder~\cite{tuckett2019tailoring, bonilla2021xzzx}.
    (c) Logical error rate $p_L$ for all $3^{d^2}$ distance $d=3$ Clifford-deformed surface codes as a function of the conditional entropy $H(L|S)$ in the code capacity setting and Pauli biased noise with $p_X=p_Y=p_Z/2\eta$, where $ p_X +p_Y +p_Z = 0.01$ and $\eta = 500$.
    Dashed lines capture the bounds in Eqs.~\eqref{eq:p_L_lower_bound} and~\eqref{eq:p_L_upper_bound}.
    Note that Eq.~\eqref{eq:p_L_lower_bound} is Fano's inequality~\cite{fano1961transmission} and Eq.~\eqref{eq:p_L_upper_bound} is the Hellman-Raviv bound~\cite{hellman1970probability}.}
    \label{fig:noise-bias}
\end{figure*}

The connection between entropy and error correction is the foundational insight in information theory~\cite{Cover2005}.
Concretely, Shannon's noisy-channel coding theorem~\cite{shannon1948mathematical} asserts that classical information can be reliably transmitted through a noisy channel at rate $R$ if and only if $R < C$, where $C$ is the channel capacity; for a symmetric binary channel with bit-flip error rate $p$, $C = 1 - H_2(p)$, where $H_2(p) = -(1-p) \log_2(1-p) - p \log_2(p)$ is the binary entropy.
This connection has been emphasized in numerous studies~\cite{hamming1950error,gilbert1952comparison,varshamov1957estimate} and also adapted to the quantum setting~\cite{calderbank1996good,bennett1996mixed,lloyd1997capacity,shor2002quantum,devetak2005private}, establishing some upper and lower bounds on the rates achievable for QEC protocols.
Notably, the hashing bound provides a lower bound on the threshold of random QEC codes, and the surface code~\cite{kitaev2003fault,dennis2002topological} performs remarkably close to this limit.
Its code capacity threshold ($\approx 10.9\%$) approaches the zero-rate hashing bound of $p \approx 11\%$, which is derived by setting $R = 1 - 2H_2(p) = 0$~\cite{dennis2002topological}.
Alternative methods based on coherent information~\cite{colmenarez2024accurate} or mappings to statistical mechanical~\cite{dennis2002topological,katzgraber2009error,bombin2012strong,Kubica2018,Chubb2021,English2025} and stochastic kinetic~\cite{sriram2026kinetics} models also provide precise threshold estimates, but are computationally expensive.
Here, we take an alternative approach, demonstrating that entropy can also be applied in a practical setting to derive an easily calculable estimate for thresholds of QEC protocols.

\textit{Threshold heuristic}.---The performance of a QEC protocol can be characterized by its logical error rate $p_L$, i.e., the probability of the encoded information being corrupted by noise.
Here, we consider stabilizer codes which detect errors, represented by the random variable $E$,
by measuring stabilizer operators.
The measurement outcome $S$ is a string of bits, each corresponding to a value of a random variable associated with the stabilizer; we refer to $S$ as the error syndrome.
Then, the uncertainty about the logical Pauli error class $L$ given the syndrome $S$ is given by the conditional entropy $H(L|S) = -\sum_{l,s} \pr(l,s)\log_2 \pr(l|s)$, where the summation is over all possible values of $L$ and $S$.
Even though $H(L|S)$ does not directly determine $p_L$, it provides a lower and an upper-bound, i.e.,
\begin{align}
H(L|S) &\le H_2(p_L) + p_L\log_2 (4^k-1),
\label{eq:p_L_lower_bound}
\\
p_L &\le \tfrac{1}{2}H(L|S),
\label{eq:p_L_upper_bound}
\end{align}
where $k$ is the number of logical qubits encoded and $p_L$ is the logical error rate under a maximum-likelihood decoder; see Appendix~\ref{app:bound_pL} for details.
We illustrate these bounds for Clifford-deformed surface codes~\cite{dua2024clifford}; see Fig.~\ref{fig:noise-bias}(a)(c).
Although the relationship between $H(L|S)$ and $p_L$ is not strictly monotonic, the strong correlation demonstrates that the conditional entropy serves as a reliable predictor of the logical error rate.
More crucially, Eqs.~\eqref{eq:p_L_lower_bound} and~\eqref{eq:p_L_upper_bound} imply $H(L|S)$ exhibits a threshold behavior at the QEC threshold $p_{th}$. For example, assuming independent and identically distributed depolarizing noise, if its strength $p>p_{th}$, then $p_L$ does not vanish as $d\to\infty$, implying that $H(L|S)$ does not vanish either; if $p<p_{th}$, then $p_L \xrightarrow{d\to\infty} 0$, implying $H(L|S) \xrightarrow{d\to\infty} 0$.\footnote{This implication requires $k\,p_L\to0$, which holds below threshold since $p_L$ decays exponentially with the distance while $k$ grows at most polynomially with the code size.}

However, calculating $H(L|S)$ is computationally expensive as it involves the probability distribution over exponentially many possible syndromes.
Nevertheless, using the chain rule, $H(L|S)$ can be written as follows
\begin{equation}
    H(L|S) = H(E) - H(S) - H(E|L,S),
\label{eq:entropy_balance}
\end{equation}
where, for a discrete random variable $X$ with probability distribution $\pr(x)$, the entropy is defined as $H(X) = -\sum_x \pr(x)\log_2 \pr(x)$, and $E$ represents the physical error configuration,
e.g., for the circuit-level model, the Pauli errors occurring at every space-time location in the circuit.
The decomposition of Eq.~\eqref{eq:entropy_balance} sheds light on the different terms contributing to the uncertainty of the logical operator.
$H(E)$ represents the total uncertainty introduced by the errors; we will call it the \textit{noise entropy}. $H(S)$ represents the information acquired by the stabilizer measurements (it can also incorporate flag measurements or erasure checks), which are designed to identify the errors; we will call it the \textit{syndrome entropy}.
Finally, $H(E|L,S)$ quantifies the entropy associated with the degeneracy of the QEC protocol, that is, the fact that different error configurations may have the same syndrome and belong to the same logical equivalence class; we will call it the \textit{degeneracy entropy}.
As entropy is non-negative, the degeneracy reduces the uncertainty on the logical error and thus contributes to reducing the logical error rate.
Indeed, a QEC protocol does not need to identify the physical error $E$ exactly, but only determine its logical equivalence class.

\begin{figure}
    \centering
    \includegraphics[width=\columnwidth]{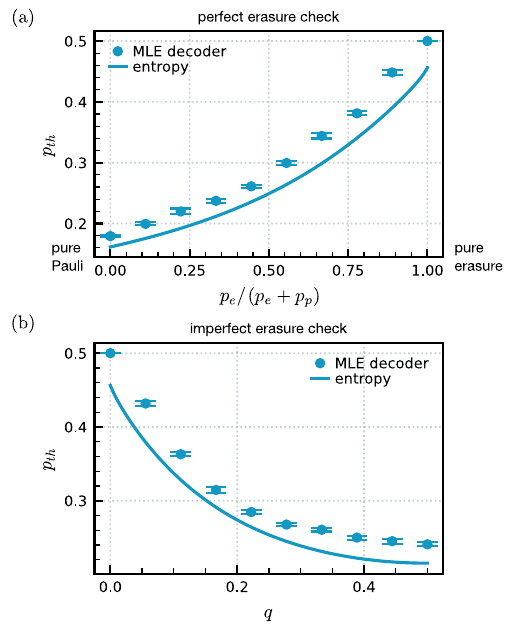}
    \caption{(a) Code capacity threshold of the surface code for a mix of erasure noise, with probability $p_e$, and Pauli noise, with probability $p_p$. (b) Surface code threshold for pure erasure noise and imperfect erasure checks. We set the false-negative probability to $\pr(F = 0 \mid \text{erasure}) = q$ and the false-positive probability to $\pr(F=1\mid \text{no erasure} ) = q$. The most-likely error decoding is performed with the Gurobi optimizer~\cite{gurobi}.}
    \label{fig:erasure_capacity}
\end{figure}

The decomposition in Eq.~\eqref{eq:entropy_balance} motivates a simple estimator of the QEC threshold, which is defined with respect to a family of QEC codes of growing size. 
For the $i$th QEC code in the family, we denote by $n_i$, $k_i$, $d_i$ and $m_i$ the number of physical and logical qubits, code distance and the number of independent stabilizer operators;
we also denote by $E_i$, $S_i$ and $L_i$ the random variables corresponding to the physical error configuration, the syndrome and the logical error class.
For Pauli noise, these variables are represented as binary random vectors of lengths $2n_i$, $m_i$ and $2k_i$, respectively.
For a family of QEC codes operating below threshold, $H(L_i|S_i) \to 0$, and we typically expect that, for sufficiently large $i$,
\begin{align}
H(E_{i+1}) - H(E_i) \le & H(S_{i+1}) + H(E_{i+1}|L_{i+1},S_{i+1}) \nonumber \\
& - H(S_i) - H(E_i|L_i,S_i).
\label{eq:threshold_criteria}
\end{align}
This corresponds to the situation where the amount of the additional syndrome information extracted by a larger QEC code exceeds the additional noise entropy, with a balance achieved at the threshold.
Usually, $H(E)$ is easy to compute as errors are assumed independent on the qubits or within the circuit; on the other hand, computing $H(S)$ and $H(E|L,S)$ requires exploring the exponentially-large space of all possible syndrome outcomes.
Remarkably, we find that comparing the noise entropy with the syndrome entropy within a unit cell of the code or the circuit, both of which are local and easy to compute, yields an accurate heuristic for estimating the threshold.
More precisely, following Eq.~\eqref{eq:threshold_criteria}, we evaluate the threshold as the physical error rate at which the change of noise entropy with code size is equal to the change of syndrome entropy with code size, both evaluated locally.

As a guiding example, we detail the construction for the code-capacity setting. Let $E=(E^1,\ldots,E^n)$ denote the physical error configuration on the $n$ qubits.
Assuming that errors on different qubits are independent and identically distributed, we have
\begin{equation}
H(E)=\sum_{i=1}^{n}H(E^i)=n \hn(E),
\end{equation}
where $\hn(E)$ is the noise entropy per qubit.
To estimate the syndrome entropy, we approximate the syndrome bits as independent. Let $S=(S^1,\ldots,S^m)$ denote the $m$-bit syndrome. For simplicity, we consider here a code whose syndrome bits all have the same marginal entropy in the large-size limit, as is the case for the surface or color code under depolarizing noise; we then have
\begin{equation}
H(S)\approx\sum_{i=1}^{m}H(S^i)=m \hn(S),
\end{equation}
where $\hn(S)$ is the marginal entropy per syndrome bit. The threshold \(p_\star\) is estimated as the value of \(p\) for which the changes of the two entropies between successive members of the family are equal, i.e.,
\begin{equation}
H(E_{i+1}) - H(E_i)=H(S_{i+1}) - H(S_i).
\label{eq:entropy_condition}
\end{equation}
Using the expressions above, denoting by $\hn_E(p)$ and $\hn_S(p)$ the noise and syndrome entropies, respectively, for an error of strength $p$, this condition becomes
\begin{equation}
(n_{i+1} - n_i) \hn_E(p_\star) = (m_{i+1} - m_i) \hn_S(p_\star).
\label{eq_hehs}
\end{equation}
For code families with $(m_{i+1} - m_i)/(n_{i+1} - n_i)\to1$ as $i\rightarrow\infty$, such as the surface code,
Eq.~\eqref{eq_hehs} reduces to
\begin{equation}
\hn_E(p_\star)=\hn_S(p_\star).
\label{eq:one_qubit_comparison}
\end{equation}
More generally, for many LDPC code families where the ratio 
$(m_{i+1} - m_i)/(n_{i+1} - n_i)$
converges, we obtain
\begin{equation}
 \hn_E(p_\star) = \lim_{i \rightarrow \infty} \frac{m_{i+1} - m_i}{n_{i+1} - n_i} \hn_S(p_\star).
\end{equation}
Thus, the estimated threshold amounts to comparing the noise entropy with the syndrome entropy per unit cell, illustrated here using a unit cell containing one qubit and one stabilizer. In the following, we generalize the notation $\widetilde H$ to denote the entropy per unit cell, rather than only per qubit or per syndrome bit.

This framework can be extended to circuit-level noise, as we will discuss later. The estimate thus relies on two assumptions:
(i) different syndrome bits are close to independent at physical error rates near the threshold,
and (ii) the degeneracy entropy $H(E|L,S)$ is small compared to the syndrome entropy.
In the End Matter, we give more details about these two assumptions using exactly computable small-scale examples to justify their validity for topological codes.

\begin{figure}
    \centering
    \includegraphics[width=\columnwidth]{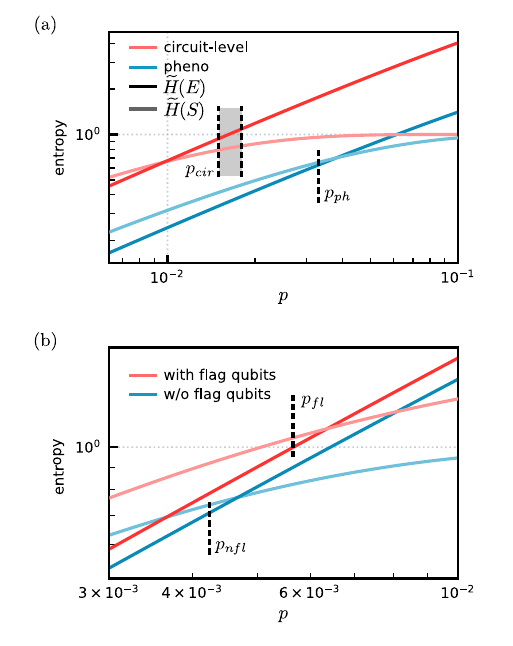}
    \caption{(a) Surface code thresholds for the phenomenological and circuit-level noise models.
    The vertical lines correspond to numerical simulations; the dark (light) colored lines correspond to the noise (syndrome) entropy. The entropies are evaluated within a unit cell of the circuit.
    For the phenomenological noise, we find the threshold of 3.8\% compared to 3.3\% from the statistical-mechanical mapping~\cite{ohno2004phase}; for circuit-level noise, we find 0.97\% compared to 1.5--1.8\% for an optimal decoder~\cite{heim2016optimal}. (b) Color code threshold with ($p_{fl}$) or without ($p_{nfl}$) the flag information given to the decoder.
    The vertical lines correspond to an approximate most-likely error decoder using the Gurobi optimizer~\cite{gurobi}.
    }
    \label{fig:circuit-level}
\end{figure}

The approach presented here is just one possible way of evaluating the threshold. A more accurate estimate of the syndrome entropy could be derived by considering larger neighborhoods, while the degeneracy entropy might be more carefully bounded. In the rest of this article, we refer to such a way of estimating the QEC threshold based on the local entropy as the \textit{entropy threshold}.

\textit{Code capacity threshold estimations}.---
First, we evaluate the threshold of the surface, XZZX and XY codes in the code capacity setting with the Pauli error rate $p = p_X + p_Y + p_Z$ and the noise bias $\eta = p_Z/(p_X + p_Y)$ with $p_X = p_Y$, where $X$ and $Y$ errors are less likely than $Z$-type errors.
As discussed above, it suffices to compare the noise and syndrome entropies per unit cell.
The noise entropy per unit cell, which for the surface code contains one data qubit and one check, alternately of $X$ and $Z$ type, is
\begin{align}
    \hn(E) &= - (1-p)\log_2(1-p) - p_X\log_2(p_X) \nonumber \\  
    &\quad - p_Y\log_2(p_Y) - p_Z\log_2(p_Z). 
\end{align}
Under the assumption of independent syndrome bits, the syndrome entropy per unit cell is 
\begin{align}
\hn(S) = \tfrac{1}{2} \left[  H_2(\pr(S_X = +1)) + H_2(\pr(S_Z = +1)) \right].  
\end{align}
It can be similarly calculated for the XZZX and XY codes. Fig.~\ref{fig:noise-bias}(b) plots the evaluated thresholds, in good agreement with numerical simulation results obtained with a tensor network decoder~\cite{tuckett2019tailoring, bonilla2021xzzx}.

The entropy balance also provides insight as to why the XZZX and XY codes are more robust against biased noise, whereas the standard surface code is not.
When increasing the noise bias toward $Z$ errors, the noise entropy decreases and all the stabilizers of the XZZX and XY codes remain sensitive.
On the other hand, only half of the stabilizers of the surface code can detect $Z$ errors, effectively reducing the syndrome entropy per unit cell by a factor of two at infinite bias.

The threshold heuristic can be adapted to the case of erasure errors.
An erasure error occurs when the location of an error is known, but its effect on the affected qubit is unknown; here, we assume that a fully depolarizing channel is applied at the locations of erasures.
The method could also be applied to biased erasures, where $Z$ errors are dominant.
Eq.~\eqref{eq:entropy_balance} can be adapted as
\begin{equation}
    H(L|S,F) = H(E|F) - H(S|F) - H(E|L,S,F),
\label{eq:erasure_balance}
\end{equation}
where $F$ denotes the information from the erasure checks. The entropy threshold then corresponds to the point where the noise entropy $H(E|F)$ is equal to the syndrome entropy $H(S|F)$, evaluated locally.
We consider the surface code in the code capacity setting, with the erasure $p_e$ and depolarization $p_p = p_X+p_Y+p_Z$ rates.
We use a model where the erasure corresponds to a heralded fully depolarizing channel applied to a qubit. Fig.~\ref{fig:erasure_capacity} shows the surface code threshold with (a) perfect and (b) imperfect erasure information.
In both cases, the entropy threshold is in good agreement with the numerical simulations. In Appendix~\ref{app:erasure}, we detail the entropy calculation for the perfect and imperfect erasure checks. 

As in the case of noise bias, the analysis of the entropy balance gives us insight into the increase of threshold under erasure noise. While the syndrome entropy stays almost constant, the noise entropy decreases with erasure noise, resulting in an increased threshold. In contrast, when the rate of false positives or negatives increases, the noise entropy increases significantly, while the syndrome entropy stays relatively constant.

\textit{Circuit-level threshold estimation.}---Thus far, we have considered only ideal stabilizer measurements, where the threshold heuristic simply compares the noise entropy with the syndrome entropy per unit cell of the code.
However, this framework can be generalized to quantum circuits through the detector formalism, as most quantum circuits inherit spatial translation invariance from the underlying code structure and often exhibit temporal translation invariance as well, since they consist of repeated rounds of stabilizer measurements. A detector is a product of stabilizer measurement outcomes that deterministically yields +1 in the absence of errors~\cite{dennis2002topological,gidney2021stim}. The associated detecting region defines the spacetime volume in which faults will trigger this detector.
To extend our entropy formalism, the syndrome entropy will now refer to the entropy associated with the detectors in the circuit, while the noise entropy will refer to all the possible error mechanisms in the circuit.
The entropy threshold can then be evaluated by comparing the noise entropy of the errors per unit cell of the circuit, with the syndrome entropy of the detectors per unit cell.
For example, in a surface code under phenomenological noise, there is one detector, one measurement error and two data qubit errors per unit cell.
In Fig.~\ref{fig:circuit-level}(a), we compare the entropy threshold with numerical simulations of the surface code under phenomenological noise and circuit-level noise, demonstrating strong agreement between our heuristic and the numerical simulations.

To further test the method beyond surface codes, we evaluate the threshold of color codes on a hexagonal lattice~\cite{bombin2006topological}. The weight-6 $X$ and $Z$-type checks of the color code are measured using two ancilla qubits prepared in a Bell state.
Each ancilla qubit interacts with three neighboring data qubits, after which the Bell pair is disentangled and measured~\cite{baireuther2019neural}.
The resulting two bits capture the stabilizer measurement outcome and the flag.
Crucially, this flag preserves the effective distance of the code during the measurement circuit but also improves the error threshold.
Fig.~\ref{fig:circuit-level}(b) compares the color code threshold, with and without providing the flag information to the decoder, against the entropy threshold.
We find that the entropy formalism correctly accounts for the increase in syndrome entropy provided by the flag qubit and the improvement in the color code threshold.

\textit{Discussion.}---The entropy heuristic provides a readily computable estimate of the QEC threshold.
We have demonstrated that by locally comparing the noise and syndrome entropies, our approach accurately predicts surface and color code thresholds for biased Pauli, erasure, phenomenological, and circuit-level noise models.
Future work may further investigate the theoretical validity of our assumptions, specifically treating the syndrome bits as independent and neglecting the degeneracy entropy.
For instance, in the End Matter, we show that the low-degeneracy assumption breaks down for the Bacon-Shor code~\cite{bacon2006operator}, which exhibits even greater degeneracy than topological codes. A rigorous analysis of these terms could potentially establish formal bounds on the accuracy of our heuristic or lead to a refined version of the heuristic itself. The procedure presented in this work should be viewed as one instantiation of the entropy-based approach, rather than a definitive implementation of the method. Ultimately, we anticipate that this framework could be leveraged to optimize QEC strategies, utilizing the entropy-based threshold as a cost function. Another potential direction is to extend the framework to non-Pauli noise. While the noise can be discretized by the stabilizer measurements, the noise entropy would then need to account for non-binary contributions. Along the same lines, in the End Matter, we discuss how the framework can be adapted to incorporate soft information from stabilizer measurements. The code used for generating the data can be found at~\cite{ruiz2026entropythreshold}.

\begin{acknowledgements}

A.K. acknowledges support from the U.S. Army Research Office (Grant No.~W911NF-23-1-0051) and IARPA (ELQ Program, Cooperative Agreement No. W911NF-23-2-0219).

\end{acknowledgements}

\bibliography{biblio}

\clearpage

\section*{End matter}

\begin{figure*}
    \centering
    \includegraphics[width=\textwidth]{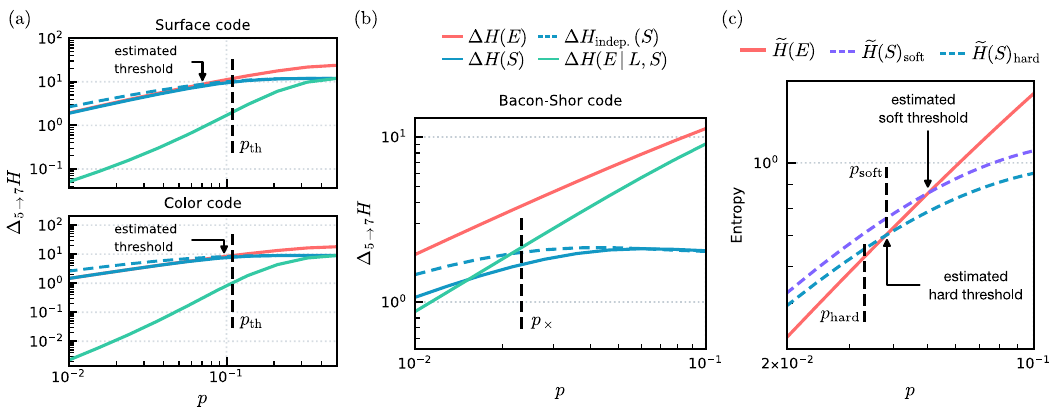}
    \caption{(a) Change of the entropic contributions for the surface and color codes between distances $d=5$ and $d=7$, in the code capacity setting and Pauli $Z$ errors of strength $p$.
    When $p \approx 50\%$, the degeneracy entropy $H(E|L,S)$ is comparable with the syndrome entropy $H(S)$, but it rapidly becomes negligible as the error rate decreases.
    We also observe that close to the threshold (dashed vertical lines), treating syndrome bits as independent gives rise to a good approximation of the syndrome entropy.
    This approximation, however, does not work for every code, as exemplified by the Bacon-Shor code in (b).
    Here, $p_\times$ is the physical error rate at which the logical error rates of Bacon-Shor codes for $d=5$ and $d=7$  cross.
    In (c), we evaluate the entropy threshold for the surface code with phenomenological noise of strength $p$ (for both qubit and measurement errors) with either hard or soft measurement. The entropy is evaluated per unit cell of the circuit. While the estimate is relatively accurate for hard measurements, it is too optimistic for soft measurements.}
    \label{fig:end_matter}
\end{figure*}

\textit{Entropy threshold hypotheses.}---As described in the main text, the threshold estimate relies on two hypotheses. The first one is the \textit{independent syndrome hypothesis}. Although stabilizer measurement outcomes are strongly correlated at low physical error rates, they become increasingly decorrelated as the physical error rate increases. Indeed, when errors are sparse, two stabilizers sharing a qubit have their measurements highly correlated. Observing one flipped stabilizer increases the probability that other stabilizers sharing qubits with it are flipped as well. In contrast, when the physical error rate is high, errors are so frequent that the value of a given stabilizer carries little information about other stabilizers sharing qubits with it. Empirically, we observe that for the surface code and the color code in the code capacity setting, at the threshold value, the following approximation holds 
\begin{equation}
    H(S) \approx \sum_i H(S^i),
\end{equation}
where $S$ is the random variable describing the syndrome and $S^i$ is the marginal random variable for its $i$th bit.

We refer to the second hypothesis we make as the \textit{low-degeneracy hypothesis}. Although codes such as the surface code are known to be highly degenerate, we empirically find that even for such codes, 
\begin{equation}
H(E_{i+1}|L_{i+1},S_{i+1}) - H(E_{i}|L_{i},S_{i}) \ll H(S_{i+1}) - H(S_{i}),
\end{equation}
where $i$ refers to the surface or the color code of distance $d = 2i+1$. This means that the dominating term contributing to reducing the error entropy in Eq.~\eqref{eq:entropy_balance} is the syndrome entropy $H(S)$. The two hypotheses are illustrated in Fig.~\ref{fig:end_matter}(a) for the surface and color codes in the code capacity setting where all quantities can be calculated exactly. Interestingly, we also see that the approximation is tighter for the color code, as it has a lower degeneracy entropy.

Combining these two hypotheses, the condition for being at the threshold sums up to Eq.~\eqref{eq:entropy_condition}. Both terms are now easily computable and intuitively, Eq.~\eqref{eq:entropy_condition} corresponds to the point where the uncertainty added by the errors is small enough to be correctly identified by the stabilizer measurements.

Even though these estimates seem to hold for topological codes, it is not the case for all codes. To understand the limit of our estimate beyond topological codes, we investigate Bacon-Shor codes~\cite{bacon2006operator}. The gauge group of the code is local, but its stabilizers are of weight $2\times \ell$ for a lattice of size $\ell$. The code has no threshold, meaning that for any physical error $p>0$, there is a logical error rate that cannot be reached~\cite{NappPreskill2013}. This is captured by our estimate, following Eq.~\eqref{eq:entropy_balance}. First, applying our method, the degeneracy entropy $H(E|L,S)$ is neglected. The change of noise entropy scales as $\Delta H(E) \propto \ell$ while the change of syndrome entropy scales as $\Delta H(S) \propto 1$, such that asymptotically we always have $\Delta H(E) > \Delta H(S)$. The method thus predicts correctly that the code has no threshold. We note that this argument does not apply to the alternative Bacon-Shor measurement schedules proposed in Refs.~\cite{gidney2023less,alam2025bacon}, which possess hierarchical and bounded-weight spacetime detectors, respectively.

However, when evaluating how the terms in Eq.~\eqref{eq:entropy_balance} vary across low-distance instances of the Bacon--Shor code and code-capacity noise, we find that the degeneracy entropy is substantially larger than for surface and color codes. Consequently, the low-degeneracy hypothesis no longer holds, as shown in Fig.~\ref{fig:end_matter}(b). This is explained by the extensive gauge degeneracy of the Bacon-Shor code, since many physical errors at different locations along rows or columns are in the same stabilizer and logical operator eigenspace, leaving substantial residual uncertainty in $E$ when $L$ and $S$ are fixed.

\textit{Soft information.}---Finally, to go beyond fully digitized noise models, we evaluate the change in the surface-code threshold when incorporating soft information from the measurements. In most quantum platforms, a measurement does not yield a binary outcome, but an analog value which contains more information. A simple model, close to homodyne measurement of superconducting qubits, considers that a qubit in the state $\ket{0}$ (or $\ket{1}$) will result in an analog value drawn from the distribution $\mathcal N (0,\sigma^2)$ (or $\mathcal N (1,\sigma^2)$). Thus, $\sigma \ll 1$ yields very high-fidelity measurement while a $\sigma$ close to 1 yields low-fidelity measurement. For every measurement, this information can be transmitted to the decoder to update the decoding graph and improve the decoding~\cite{pattison2021improved}. In Fig.~\ref{fig:end_matter}(c), we compare the improvement in threshold predicted by the entropy threshold with the one simulated by updating the decoding graph and using matching to predict the errors. In Appendix~\ref{app:soft}, we detail how the entropy threshold framework can be adapted to soft measurement. In agreement with the simulations, the entropy method indeed predicts an increase of the threshold with soft measurement. We also observe that the entropy threshold predicts a higher value than the simulations. However, we cannot determine whether this discrepancy is due to the suboptimality of the matching graph update procedure or to the entropy threshold being overly optimistic due to the neglected terms.

\clearpage

\appendix

\onecolumngrid

\section{Logical error rate bound}
\label{app:bound_pL}

In this appendix, we show that
\begin{align}
    H(L|S) &\le H_2(p_L) + p_L\log_2 (4^{k}-1), \label{eq:fano} \\
    p_L &\le H(L|S)/2. \label{eq:hellman}
\end{align}
We recall that $L$ is the random variable of the logical class of the error,
with values in $\mathcal{L}=\{I,X,Y,Z\}^{k}$, $S$ is the measured syndrome,
$p_L$ is the logical error rate under a maximum-likelihood decoder and $k$ is the number of encoded logical qubits.

\begin{proof}
    Eq.~\eqref{eq:fano} corresponds to Fano's inequality~\cite{fano1961transmission} and Eq.~\eqref{eq:hellman} is the Hellman-Raviv bound~\cite{hellman1970probability}, that we rederive here. We denote by $F$ the binary random variable that is equal to one when the
    maximum-likelihood decoder fails and zero otherwise. We thus have
    $\pr(F=1)=p_L$. Since $F$ is determined by $L$ and $S$, the chain rule gives
    \begin{align}
        H(L|S) = H(L,F|S) = H(F|S)+H(L|F,S).
    \end{align}
    The first term satisfies
    \begin{align}
        H(F|S) \le H(F) = H_2(p_L).
    \end{align}
    For the second term,
    \begin{align}
        H(L|F,S)
        =&(1-p_L)H(L|F=0,S)+p_LH(L|F=1,S).
    \end{align}
    If $F=0$, the logical class is the one selected by the decoder, and hence
    $H(L|F=0,S)=0$. If $F=1$, the logical class is one of the $4^{k}-1$ remaining
    classes, and hence $H(L|F=1,S)\le\log_2 (4^{k}-1)$. Therefore,
    \begin{align}
        H(L|F,S)
        &\le p_L\log_2 (4^{k}-1).
    \end{align}
    Combining these results gives
    \begin{align}
        H(L|S) \le H_2(p_L)+p_L\log_2 (4^{k}-1).
    \end{align}

    For the second inequality, by definition we have
    \begin{align}
        p_L
        &=\sum_s \pr(S=s)\pr(F=1|S=s).
    \end{align}
    We denote $q_s=\pr(F=1|S=s)$. We have
    \begin{align}
        H(L|S=s) = H(F|S=s)+H(L|F,S=s)
        \ge H(F|S=s) \ge H_2(q_s).
    \end{align}
    If $q_s\le 1/2$, since $H_2$ is concave and $H_2(0) = 0$ and $H_2(1/2) = 1$, we have $H_2(q_s) \ge 2q_s$ and it follows that $H(L|S=s)\ge 2q_s$. If $q_s\ge 1/2$, we have
    \begin{align}
        H(L|S=s)
        =&-\sum_{\ell\in\mathcal{L}}
        \pr(L=\ell|S=s) \log_2 \pr(L=\ell|S=s) \\
        \ge&-\sum_{\ell\in\mathcal{L}}
        \pr(L=\ell|S=s)\log_2(1-q_s) 
        \ge -\log_2(1-q_s) \ge 2q_s,
    \end{align}
    where we used $\pr(L=\ell|S=s)\le 1-q_s$ for every $\ell\in\mathcal{L}$, since the maximum-likelihood decoder selects a most probable logical class. For the last inequality, we used the fact that $-\log_2(1-q_s)\ge 2q_s$ for all $q_s\in[1/2,1)$. Thus, in both cases,
    \begin{align}
        q_s = \pr(F=1|S=s)
        &\le H(L|S=s)/2.
    \end{align}
    Substituting this bound into the expression for $p_L$ gives
    \begin{align}
        p_L
        &\le\frac{1}{2}\sum_s \pr(S=s)
        H(L|S=s) =H(L|S)/2.
    \end{align}
\end{proof}

\section{Entropy calculation with erasure qubits}
\label{app:erasure}

In this appendix, we evaluate the noise entropy and the syndrome entropy of erasure qubits. We first consider the case of perfect erasure checks for the surface
code in a code capacity setting. We recall that $p_e$ is the erasure
probability and $p_p$ is the Pauli probability. The decoder is informed
if there was an erasure with a flag $F$, where $F=1$ if an erasure is detected, and $F=0$ otherwise.
Following Eq.~\eqref{eq:one_qubit_comparison}, throughout this Appendix, all entropies are evaluated for a unit cell of the code, i.e., for a single qubit or stabilizer. The noise entropy of one qubit is given by
\begin{align}
    \hn(E|F)
    &=\hn(E|F=0)\pr(F=0)
    +\hn(E|F=1)\pr(F=1)
    \\
    &=\left[ -(1-p_p)\log_2(1-p_p)
    -p_p\log_2\left(\frac{p_p}{3}\right) \right](1-p_e)+2p_e,
\end{align}
where we used $\hn(E|F=1)=2$, since an erased qubit is fully depolarized
and thus requires 2 bits of entropy to specify the corresponding Pauli
error among $\{I,X,Y,Z\}$. Similarly, we can calculate the syndrome
entropy. We denote by $F_i$, $i\in\{1,2,3,4\}$, the erasure flags of
the four qubits involved in the stabilizer, and write
$\mathbf F=(F_1,F_2,F_3,F_4)$. The exponent 4 appearing in the following expressions comes from the weight-4 stabilizers of the surface code. The same reasoning applies to stabilizers of arbitrary weight $w$. We then have
\begin{align}
    \hn(S|\mathbf F)
    =&\hn(S|\mathbf F=\mathbf 0)
      \pr(\mathbf F=\mathbf 0)
    +\hn(S|\mathbf F\neq\mathbf 0)
      \pr(\mathbf F\neq\mathbf 0)
    \\
    =&H_2(\pr(S=+1|\text{no erasure}))(1-p_e)^4 +H_2(0.5)\left[1-(1-p_e)^4\right],
\end{align}
where we used $\hn(S|\mathbf{F = 0}) = H_2(\pr(S=+1|\text{no erasure}))$ as $S$ is a binary variable. $\pr(S=+1|\text{no erasure})$ is the probability that the stabilizer outcome is $+1$ under Pauli noise only and under depolarizing noise is equal to  
\begin{equation}
\pr(S = +1|\text{no erasure})=\frac{1}{2}\left[1+\left(1-\frac{4p_p}{3}\right)^4\right].
\end{equation}
When at least one qubit participating in the check is erased, we instead have $\pr(S=+1|\text{at least one erasure})=0.5$.

The derivation can be adapted to imperfect erasure checks. We define
the false-negative and false-positive probabilities as
\begin{align}
    q_-&=\pr(F=0|\text{erasure}) \\
    q_+&=\pr(F=1|\text{no erasure}),
\end{align}
respectively. The noise entropy of one qubit is given by
\begin{align}
    \hn(E|F)
    &=\hn(E|F=0)\pr(F=0)
     +\hn(E|F=1)\pr(F=1)
    \\
    &=\hn(E|F=0)
      \left[(1-p_e)(1-q_+)+p_eq_-\right] +\hn(E|F=1)
      \left[p_e(1-q_-)+(1-p_e)q_+\right],
\end{align}
with
\begin{align}
    \hn(E|F=f) =& -\sum_{P \in \{I,X,Y,Z\}}\pr(E=P|F=f) \log_2 \pr(E=P|F=f).
\end{align}
The individual probabilities can be calculated as follows:
\begin{align}
    \pr(E=I|F=0) =&
    \pr(E=I|F=0,\text{erasure})
       \pr(\text{erasure}|F=0)\nonumber \\
    &+\pr(E=I|F=0,\text{no erasure})
       \pr(\text{no erasure}|F=0) \\
    =&\frac{1}{4}
       \frac{q_-p_e}
       {(1-p_e)(1-q_+)+p_eq_-}
    +(1-p_p)
       \frac{(1-q_+)(1-p_e)}
       {(1-p_e)(1-q_+)+p_eq_-},
\end{align}
and
\begin{align}
    \pr(E=X|F=0)
      =&\pr(E=Y|F=0)
    =\pr(E=Z|F=0)
    \\
    =&\frac{1}{4}
       \frac{q_-p_e}
       {(1-p_e)(1-q_+)+p_eq_-}
    +\frac{p_p}{3}
       \frac{(1-q_+)(1-p_e)}
       {(1-p_e)(1-q_+)+p_eq_-}.
\end{align}
Similarly,
\begin{align}
    \pr(E=I|F=1)
    =&\pr(E=I|F=1,\text{erasure})
       \pr(\text{erasure}|F=1)
    \nonumber\\
    &+\pr(E=I|F=1,\text{no erasure})
       \pr(\text{no erasure}|F=1)
    \\
    =&\frac{1}{4}
       \frac{(1-q_-)p_e}
       {(1-p_e)q_+ +p_e(1-q_-)}
    +(1-p_p)
       \frac{q_+(1-p_e)}
       {(1-p_e)q_+ +p_e(1-q_-)},
\end{align}
and
\begin{align}
    \pr(E=X|F=1)
      =&\pr(E=Y|F=1)
    =\pr(E=Z|F=1)
    \\
    &=\frac{1}{4}
       \frac{(1-q_-)p_e}
       {(1-p_e)q_+ +p_e(1-q_-)}
    +\frac{p_p}{3}
       \frac{q_+(1-p_e)}
       {(1-p_e)q_+ +p_e(1-q_-)}.
\end{align}

The syndrome entropy can be calculated as follows using again the
notation $F_i$, $i\in\{1,2,3,4\}$, for the erasure flags of the four
qubits involved in the stabilizer. We denote the actual erasure pattern
by $\mathbf R=(R_1,R_2,R_3,R_4)$, where $R_i=1$ if qubit $i$ is
erased and $R_i=0$ otherwise. We have
\begin{align}
    \hn(S|\mathbf F)
    &=\sum_{\mathbf f\in\mathbb F_2^4}
      \hn(S|\mathbf F=\mathbf f)
      \pr(\mathbf F=\mathbf f)
    =\sum_{\mathbf f\in\mathbb F_2^4}
      H_2\left(
      \pr(S=+1|\mathbf F=\mathbf f)
      \right)
      \pr(\mathbf F=\mathbf f).
\end{align}

The individual probabilities can be calculated as follows. For fixed
erasure check outcomes $\mathbf f$,
\begin{align}
    \pr(S=+1|\mathbf F=\mathbf f)
    &=
      \sum_{\mathbf r\in\mathbb F_2^4}
      \pr(S=+1|\mathbf F=\mathbf f,\mathbf R=\mathbf r)
      \pr(\mathbf R=\mathbf r|\mathbf F=\mathbf f)
    \\
    &=
      \sum_{\substack{\mathbf r\in\mathbb F_2^4\\|\mathbf r|>0}}
      \frac{1}{2}
      \pr(\mathbf R=\mathbf r|\mathbf F=\mathbf f)
      +\pr(S=+1|\mathbf R=\mathbf 0)
       \pr(\mathbf R=\mathbf 0|\mathbf F=\mathbf f)
    \\
    &=
      \frac{1}{2}
      +\left[
       \pr(S=+1|\mathbf R=\mathbf 0)-\frac{1}{2}
       \right]
       \pr(\mathbf R=\mathbf 0|\mathbf F=\mathbf f),
\end{align}
where $|\mathbf r|$ denotes the Hamming weight of the erasure pattern.
In the second equality, we used the fact that the probability of the stabilizer outcome being $+1$
is equal to $1/2$ whenever at least one of its qubits is erased.

For compactness, we define
\begin{align}
    a_0
    &:=\pr(F_i=0)
      =(1-p_e)(1-q_+)+p_eq_-,
    \\
    a_1
    &:=\pr(F_i=1)
      =(1-p_e)q_+ +p_e(1-q_-),
\end{align}
and
\begin{align}
    \eta_0
    &:=\pr(R_i=0|F_i=0)
      =\frac{(1-p_e)(1-q_+)}
      {(1-p_e)(1-q_+)+p_eq_-},
    \\
    \eta_1
    &:=\pr(R_i=0|F_i=1)
      =\frac{(1-p_e)q_+}
      {(1-p_e)q_+ +p_e(1-q_-)}.
\end{align}
For an erasure-check outcome pattern $\mathbf f$ of Hamming weight
$k=|\mathbf f|$, we have
\begin{equation}
    \pr(\mathbf R=\mathbf 0|\mathbf F=\mathbf f)
    =\eta_1^k\eta_0^{4-k}.
\end{equation}
Consequently, the probability of the $+1$ stabilizer outcome depends
only on the number $k$ of positive flags and is given by
\begin{align}
    &\pr(S=+1\mid |\mathbf F|=k) = \frac{1}{2}
     +\left(\pr(S=+1|\mathbf R=\mathbf 0)-\frac{1}{2}\right)
      \eta_1^k\eta_0^{4-k},
\end{align}
where $\pr(S=+1|\mathbf R=\mathbf 0)$ is simply the probability of the $+1$ outcome under Pauli noise only. 

For each specific erasure-check outcome pattern $\mathbf f$ of weight $k$,
we have
\begin{equation}
    \pr(\mathbf F=\mathbf f)=a_1^ka_0^{4-k}.
\end{equation}
There are $\binom{4}{k}$ such patterns. The syndrome entropy is
therefore
\begin{align}
    \hn(S|\mathbf F)
    &=
      \sum_{k=0}^{4}
      \binom{4}{k}
      a_1^ka_0^{4-k}
      H_2\Bigg[
      \frac{1}{2}
      +\left(\pr(S=+1|\mathbf R=\mathbf 0)-\frac{1}{2}\right)
       \eta_1^k
       \eta_0^{4-k}
      \Bigg].
\end{align}

\section{Soft information measurement}
\label{app:soft}

During syndrome extraction, the analog signal associated with each measurement is usually converted into a binary outcome. This discretization discards information about the reliability of the measurement, as analog signals close to the decision boundary are less reliable than those far from it.

With soft measurement information, instead of just giving to the decoder the binary value of the detectors, we also give the confidence of the various measurements.
This additional information is then used by the decoder to improve its accuracy. To take this into account, the random variable associated with the errors $E$, as previously, is defined as the set of data errors and binary measurement errors, where a measurement error occurs when the analog signal falls on the wrong side of the decision boundary. Meanwhile, $S$ now contains both the hard detector outcomes and the reliability information associated with each measurement.

Considering again Eq.~\eqref{eq:entropy_balance}, we have
\begin{align}
    H(L|S) &= H(L,S) - H(S) = H(E,L,S) - H(E|L,S) - H(S) = H(E,S) - H(E|L,S) - H(S) \\
    &= H(E) + H(S|E) - H(S) - H(E|L,S) = H(E) - I(E;S) - H(E|L,S).
\end{align}
We used the fact that $L$ is determined by $E$ and $S$.
Entropies involving $S$ are differential entropies as they apply to continuous variables.
We see that the equivalent quantity to $H(S)$ is now $I(E;S)=H(S)-H(S|E)$, where $H(S|E)\ne 0$ as $S$ contains analog measurement results while $E$ does not.

We consider the surface code and phenomenological noise model with the qubit $p$ and measurement $q$ error rates.
There are one detector, two data errors and one measurement error per unit cell.
A detector $D$ involves four data qubit errors and two measurement errors, thus, for hard measurements, its entropy is given by
\begin{equation}
    \hn(D)
    =
    H_2\left(
    \frac{1-(1-2q)^2(1-2p)^4}{2}
    \right).
\end{equation}
The noise entropy per unit cell is given by
\begin{equation}
    \hn(E)=2H_2(p)+H_2(q).
\end{equation}

In the soft measurement case, the noise entropy does not change, but the syndrome entropy increases, as the decoder is now also informed of the reliability of the measurements. We denote by $Q_1$ and $Q_2$ the random variables associated with the error probabilities of the two measurements involved in $D$. $Q_1$ and $Q_2$ have values in $(0,1/2]$, depending on how close to the decision boundary is the analog measurement. The information contained in the hard detector and its two analog confidences can be decomposed using the chain rule as follows
\begin{equation}
    I(E;D,Q_1,Q_2)
    =
    I(E;Q_1,Q_2)
    +I(E;D|Q_1,Q_2).
\end{equation}
The second term is given by
\begin{align}
    I(E;D|Q_1,Q_2)
    &=
    \hn(D|Q_1,Q_2)
    -\hn(D|E,Q_1,Q_2).
\end{align}
Since $E$ contains all hard data and measurement errors, it determines $D$, and therefore
\begin{equation}
    \hn(D|E,Q_1,Q_2)=0.
\end{equation}
We thus have
\begin{equation}
    I(E;D|Q_1,Q_2)
    =
    \hn(D|Q_1,Q_2).
\end{equation}
The first term in the chain rule contains the analog information. $E$ encapsulates data errors and measurement errors, but $Q_1$ and $Q_2$ only depend on the measurement errors contained in $E$ whose variables we denote $M_1$ and $M_2$. Thus, we have
\begin{equation}
    I(E;Q_1,Q_2)
    =
    I(M_1;Q_1)+I(M_2;Q_2).
\end{equation}
Each detector involves two analog confidence values, with each confidence shared between two consecutive detectors, resulting in one confidence value per unit cell. Therefore, the analog contribution per unit cell is
\begin{align}
    \frac{1}{2}I(M_1;Q_1)
    +\frac{1}{2}I(M_2;Q_2)
    \nonumber =I(M;Q),
\end{align}
where the last equality follows from the fact that the two measurement channels are identical (we thus drop the indexes 1 and 2). We thus define the syndrome information per unit cell as
\begin{equation}
\widetilde{\mathcal I}_{\mathrm{cell}} := \widetilde H(D|Q_1,Q_2) + I(M;Q).
\end{equation}
As $Q$ is a continuous variable, we have
\begin{align}
    I(M;Q)= h(Q)-h(Q|M)
\end{align}
where $h$ is the differential entropy defined as
\begin{equation}
    h(X)
    =
    -\int_{x\in\chi} f(x)\log_2 f(x)\,dx
\end{equation}
where $f$ is the probability density function of a continuous random variable $X$ on $\chi$.

Combining these two contributions, we obtain
\begin{align}
    \widetilde{\mathcal I}_{\mathrm{cell}}=\hn(D|Q_1,Q_2)+h(Q)-h(Q|M).
\end{align}
The first term can be computed as
\begin{align}
    \hn(D|Q_1,Q_2)
    &=
    \int_0^{1/2}\!\!\int_0^{1/2}
    f(q_1)f(q_2)
    H_2\left[\pr(D=1|q_1,q_2)\right]
    \,dq_1dq_2.
\end{align}
We model the measurement as normally distributed random variables $X\sim\mathcal N(0/1,\sigma^2)$ with mean $\mu = 0,1$ and standard deviation
\begin{equation}
    \sigma
    =
    \frac{1}{2\Phi^{-1}(1-q)},
\end{equation}
where $\Phi$ is the standard normal cumulative distribution function. Thus, the associated hard measurement error probability is $q$.
We then have
\begin{equation}
    Q
    =
    \frac{1}
    {1+\exp\left(\frac{|0.5-X|}{\sigma^2}\right)},
\end{equation}
with $0<Q\leq 1/2$, whose probability density function is given by
\begin{align}
f(\widetilde q)
&=
\frac{\sigma}{\sqrt{2\pi}}\,
\frac{1}{\widetilde q(1-\widetilde q)}
\Bigg[
\exp\left(
-\frac{
\left(
\sigma^2\ln\left(\frac{1-\widetilde q}{\widetilde q}\right)
-\frac12
\right)^2
}{2\sigma^2}
\right)
+
\exp\left(
-\frac{
\left(
\sigma^2\ln\left(\frac{1-\widetilde q}{\widetilde q}\right)
+\frac12
\right)^2
}{2\sigma^2}
\right)
\Bigg].
\end{align}
This density is defined for $0<\widetilde q\leq 1/2$. This allows us to compute $h(Q)$ and using
\begin{equation}
    \pr(D=1|q_1,q_2)
    =
    \frac{
    1-(1-2p)^4(1-2q_1)(1-2q_2)
    }{2}
\end{equation}
we can compute $\hn(D|Q_1,Q_2)$.

Finally,
\begin{equation}
    h(Q|M)
    =
    qh(Q|M=1)
    +(1-q)h(Q|M=0)
\end{equation}
with
\begin{align}
h(Q|M=m)
=
-\int_0^{1/2}
f(\tilde q|M=m)
\log_2\left[f(\tilde q|M=m)\right]\,d\tilde q,
\end{align}
\begin{align}
f(\widetilde q|M=0)
&=
\frac{\sigma}{\sqrt{2\pi}}\,
\frac{1}
{\widetilde q(1-\widetilde q)(1-q)}
\exp\left(
-\frac{
\left(
\sigma^2\ln\left(\frac{1-\widetilde q}{\widetilde q}\right)
-\frac12
\right)^2
}{2\sigma^2}
\right),
\end{align}
and
\begin{align}
f(\widetilde q|M=1)
&=
\frac{\sigma}{\sqrt{2\pi}}\,
\frac{1}
{\widetilde q(1-\widetilde q)q}
\exp\left(
-\frac{
\left(
\sigma^2\ln\left(\frac{1-\widetilde q}{\widetilde q}\right)
+\frac12
\right)^2
}{2\sigma^2}
\right).
\end{align}
This completes the calculation of $\widetilde{\mathcal I}_{\mathrm{cell}}$, which can be substituted for $\hn(S)$ in our heuristic.

\end{document}